\documentclass{article}
\usepackage{graphicx}
\usepackage{amsmath}
\begin{document}
\title{The generalized Roche model}
\author{{Zakir F. Seidov} \thanks{%
e-mail:seidov@bgumail.bgu.ac.il} \\ College of Judea and
Samaria,Ariel 44837, Israel} \maketitle \begin{abstract} Some
exact analytical formulas are presented for the generalized Roche
model of rotating star. The gravitational field of the central
core is described by the model of two equal-mass point centers
placed symmetrically at rotation axis with the pure imaginary $z$
coordinates. The all basic parameters of the critical figure of
the rotating massless envelope are presented in analytical form.
The existence of the \emph{concave} form of the uniformly rotating
liquid is shown for a large enough angular velocity of the
rotation.\end{abstract}
\section{Introduction}
The classic Roche model describes the equilibrium figure of the
rotating massless liquid (gas) envelope in the gravitational field
of the point mass [1,2]. It is widely used in investigation of the
structure of the rotating star with a large enough value of the
effective polytropic index, when there is a strong concentration
of a matter to the star's center and one may ignore the
self-gravitation of the outer layers of the star as compared with
the gravitation of the star's central core. In the opposite limit
of rotation of the incompressible liquid with constant density,
the Maclarain ellipsoidal models are used [1,2].\\ The next
natural generalization of the classic Roche model would be the
taking into account the effects of the non-spherical gravitational
field of the star's core, still neglecting the self-gravitation of
the (uniformly) rotating envelope. In this note we consider one of
the possible approaches to the problem [3].
\section{Basic Equations}
\subsection{Gravitational potential of two fixed centers}
We use a common set of rectangular coordinates: $x$- and $y$-axes
are in the equatorial plane, $z$-axis coincides with the axis of
rotation, the center of coordinate system $(0,0,0)$ coincides with
the star's center.\\ Let two fixed points with equal masses
$m_{1}=m_{2}=M/2$ be placed at $z$-axis such that coordinates of
the '$centers$' are $\left( 0,0,z_{1}\right) $ and $\left(
0,0,z_{2}\right)$, where $z_{1}=i\,c,\;z_{2}=-i\,c,\;i^{2}=-1$.
\\Then gravitational potential has the axis of symmetry, $z$-axis,
(axis of rotation), and the plane of symmetry, $z=0$ plane,
(equatorial plane). For
our problem it is sufficient to consider the gravitational potential in $%
\left( x,z\right)$-plane (meridional plane) \begin{equation}
V(x,z)=\frac{G\, M}{2}\,(\frac{1}{r_{1}}+\frac{1}{r_{2}}),
\label{Vxz}\end{equation}\begin{equation}
r_{1}^{2}=x^{2}+(z-i\,c)^{2},\ r_{2}^{2}=x^{2}+(z+i\,c)^{2},
\label{r1r2}\end{equation} where $G$ is the gravitational
constant, $M$ is the star mass, $r_{1}$ and $r_{2}$ are distances
from the point $(x,z)$ to fixed centers, $c$ is the (positive)
constant of the two fixed centers problem describing the deviation
of the gravitational potential from the spherical symmetry.
Hereafter all formulas reduce to the classical Roche model case at
the limit $c\rightarrow 0$.
\subsection{Reality of gravitational potential}
Gravitational potential (\ref{Vxz},\ref{r1r2}) is of course real
(while $r_{1}$ and $r_{2}$ not), and in order to show it we
introduce, instead of rectangular coordinates $\left( x,z\right)
$, two real positive variables
 $\lambda $ and $\mu $ defined by the following relations:
\begin{equation}
r_{1}=c\,(\lambda -i\,\mu ),\ \ r_{2}=c\,(\lambda +i\,\mu ),  \label{rrlamu}
\end{equation} \begin{equation}
\lambda =\frac{r_{1}+r_{2}}{2\ c},\ \ \mu =\frac{r_{2}-r_{1}}{2\
c}, \label{lamu}\end{equation}\begin{equation}
x^{2}=c^{2}\,(1+\lambda ^{2})\,(1-\mu ^{2}),\ \ z^{2}=c^{2}\
\lambda ^{2}\ \mu ^{2}. \label{xzlamu} \end{equation} From
Equations (\ref{Vxz})--(\ref{xzlamu}) we get \begin{equation}
V(\lambda ,\mu )=\frac{G\,M}{c}\frac{\lambda }{\lambda ^{2}+\mu
^{2}}, \label{Vlamu} \end{equation} which is real, QED. \\Note
that $(\lambda ,\mu )$ coordinates are sometimes called
(\emph{oblate} or \emph{prolate}) confocal ellipsoidal
coordinates, or simply ellipsoidal coordinates, or even elliptic
coordinates, see e.g. [4,5]. Also notes that authors use different
notations for these coordinates. At $(x,z)-$plane, the curves
$\lambda $=const are confocal ellipses with focuses at points
$(x=0,\,z=+c)$ and $(x=0,\,z=-c)$ while the curves $\mu $=const
being normal to curves $\lambda $=const are hyperbolas.
\subsection{Total potential}
Potential of rotational forces is \begin{equation}
U(x)=\frac{1}{2}\,\omega ^{2}\,x^{2},  \label{Vrot} \end{equation}
where $\omega=$const is the angular velocity of rotation. The
total potential is a sum of gravitational and rotational
potentials
\begin{equation} \Phi (x,z)=V(x,z)+U(x).  \label{Phi} \end{equation}
Again we note that the total potential has an axis and a plane of
symmetry though we consider only $(x,z)-$plane (meridian plane)
and more specifically, first quadrant of $(x,z)-$plane with
\emph{positive} $x$ and $z$.
\section{Arbitrary rotation in the classic Roche problem}
\begin{figure}[tbh]\centerline{
\includegraphics{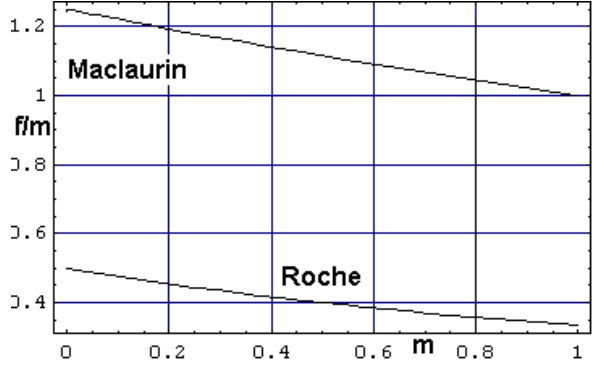}}
\caption{Ellipticity of rotating fluid in Maclaurin and Roche
models: abscissae is $m$, centrifugal-to-gravitational force ratio
at equator of the figure, ordinate is $f/m$, where $f=1-b/a$ is
ellipticity, and $a$ and $b$ are equatorial and polar radii.}
\label{elliptMR}\end{figure}

The form of the equilibrium rotating massless envelope in the CRM
depends on degree of rotation: non-rotating envelope has spherical
form, for slow rotation the figure is spheroid (figure of rotation
with meridional section as ellipse), for larger rotation the
figure is more and more oblate (and not spheroid), until the
critical figure is attained, when at the equator of the figure,
the centrifugal force is equal to gravitational force. Note that
critical figure has a \emph{cusp} at equator.
 For arbitrary value of angular velocity $\omega $ the total
 potential at the equator (where $r=x=a$) is \begin{equation}
\Phi _{c}=\frac{GM}{a}+\frac{1}{2}\omega ^{2}a^{2}.
\label{PhicCRM} \end{equation} The equilibrium figure is defined
by the equation \begin{equation} \frac{GM}{r}+\frac{1}{2}\omega
^{2}x^{2}=\frac{GM}{a}+\frac{1}{2}\omega ^{2}a^{2},
\label{eqfigCRM} \end{equation} from here we find the polar radius
($x=0,r=z=b$) \begin{equation}
\frac{GM}{b}=\frac{GM}{a}+\frac{1}{2}\omega ^{2}a^{2}
\label{polradCRM};\end{equation}
 for the ellipticity $f=1-a/b$ of the equilibrium figure
 and centrifugal-to-gravitational force ratio
at equator $m=\omega ^{2}a/(GM/a^{2})$ we have the general
expression \begin{equation} \frac{f}{m}=\frac{1}{2+m}.
\label{ftomCRM}\end{equation}
 We see that, in the CRM, the ratio $f/m$ varies from maximal
value of $1/2$ (for slow rotation, $m\ll 1$), to minimal value of
$1/3$ (for maximal rotation, $m=1$). Note that sometimes authors
use in $m$ centrifugal force at equator and \emph{mean}
gravitational force which is OK in the first approximation in
$\omega^{2}$ but has no much sense at strong rotation. \\For
comparison, for Maclaurin spheroids, the $f/m$ ratio varies from
5/4 to 1 for $m$ varying from 0 to 1, see Fig. \ref{elliptMR}.
\section{Critical figure in the GRM: \\Polar and equatorial radii}
Let us write down the gravitational potential (1,2) and the total
potential (\ref{Phi}) on $x$-axis, where $z=0,\ \
r_{1}=r_{2}=\sqrt{x^{2}-c^{2}},\ \ x>c>0$
\begin{equation}
V(x)=\frac{G\,M}{\sqrt{x^{2}-c^{2}}},\ \ \Phi (x)=V(x)+U(x).  \label{VPX}
\end{equation}
The critical surface is defined by a condition of equality of centrifugal
and gravitational forces, -$\partial V(x)/\partial x=\partial U(x)/\partial
x $, or by a minimum of the total potential, $\partial \Phi (x)/\partial x=0$%
, at some $x=a$. This gives the equation for the equatorial radius $a$:
\begin{equation}
\frac{G\,M}{\left( a^{2}-c^{2}\right) ^{3/2}}=\omega ^{2}.  \label{conda}
\end{equation} \subsection{Equatorial radius}
From Equation (\ref{conda}), the equatorial radius of the critical
surface is \begin{equation} a=\sqrt{\left( \frac{G\,M}{\omega
^{2}}\right) ^{2/3}+c^{2}}.  \label{eqrad} \end{equation}
We note that the equatorial radius of the rotating star with the same mass $%
M $ and angular velocity $\omega $ is larger than the equatorial radius $%
a_{0}$ in the classical Roche model \begin{equation} a_{0}=\left(
\frac{G\,M}{\omega ^{2}}\right) ^{1/3}.
\label{clasa}\end{equation} \subsection{Polar radius} From
equations (\ref{Vrot})-(\ref{eqrad}), we write the total potential
at the critical surface as \begin{equation} \Phi
_{0}=\frac{G\,M}{2}\frac{3\,a^{2}-2\,c^{2}}{\sqrt{\left(
a^{2}-c^{2}\right) ^{3}}}=\frac{\omega ^{2}}{2}\left(
3\,a^{2}-2\,c^{2}\right) . \label{criPhi}\end{equation} On
rotational, $z$, axis, where $x=0,\ \ r_{1}=z-c\,i,\ \
r_{2}=z+c\,i,\ \ z>c>0$ , we have only the gravitational potential
(because the rotational potential vanishes at $z$-axis)
\begin{equation} V(z)=G\ M\frac{z}{z^{2}+c^{2}}.  \label{Vz}
\end{equation} The polar radius of the critical figure is defined
by the condition $\Phi_{0}=V(z)$, and from equations
(\ref{criPhi}) and (\ref{Vz}), we find the polar radius of the
critical figure \begin{equation} b=A+\sqrt{A^{2}-c^{2}},\ \
A=\frac{(a^{2}-c^{2})^{3/2}}{3a^{2}-2c^{2}}. \label{polrad}
\end{equation} Note that we choose plus sign before radical in (\ref{polrad}) such that $%
b>c $, also we note that usually the value of the constant $c$ of
the two fixed centers model is much less than dimensions of the
figure (star or planet).\\ From (\ref{conda}) we have unequality
for the equatorial radius $a>c$, additionally from condition
$b>c$, ($A>c$), and (\ref{polrad}), we get the lower boundary for
the equatorial radius \begin{equation}
a>a_{1},\ \ \frac{a_{1}}{c}=\sqrt{4+{\left( \frac{73-\sqrt{5}}{2}\right) }^{1/3}+{%
\left( \frac{73+\sqrt{5}}{2}\right) }^{1/3}}. \label{aa1c}
\end{equation} Numerically, $a_{1}/c=3.26092$ [5].
\subsection{Two limits} At limit of small deviations from the classical
Roche model, $a\gg c$, $b\gg c$, we have for equatorial, $a$, and
polar, $b$, radii and their ratio $b/a$ \begin{equation}
a=a_0\,\left (1+{\frac{1}{2}}\, {\frac{c^2}{a_0^2}} \right ),\ \
a_0=\left ({ \frac{G\,M}{\omega ^2}}\right )^{1/3}, \end{equation}
\begin{equation}
b=b_0\,\left (1-{\frac{31}{12}}\,{\frac{c^2}{a_0^2}} \right),\ \
b_0={\frac{2 }{3}}\,a_0,\end{equation}\begin{equation}
{\frac{b}{a}}=\frac{2}{3}-\frac{37}{18}\,\left( \frac{c}{a_{0}}\right) ^{2}-%
\frac{1025}{216}\,\left( \frac{c}{a_{0}}\right) ^{4}.  \label{ba23}
\end{equation}We note that comparing with CRM, in GRM, the
equatorial radius $a$ is larger, while the polar radius $b$ and
the polar-to-equatorial radius ratio $ b/a$ is less. That is in
GRM, the critical figure of rotating envelope is more flatten
(oblate).\\ In another limit, at the equatorial radius $a$ close
to $a_{1}$ we have \begin{equation}
\frac{a}{c}=\frac{a_{1}}{c}+\delta,\ \ 0<\delta\leq
\frac{a_{1}}{c};\label{acdelta}\end{equation}
\begin{equation}\frac{b}{c}= 1+\left[
\frac{6a_{1}^{3}c\delta}{(a_{1}^{2}-c^{2})( 3 a_{1}^{2}-2
c^{2})}\right]^{1/2};\label{bcdelta}\end{equation}
\begin{equation}
\frac{b}{a}=\frac{c}{a_{1}}\left(1-\frac{c}{a_{1}}\delta +[1+\frac{%
6a_{1}^{3}\delta
c}{(a_{1}^{2}-c^{2})(3a_{1}^{2}-2c^{2})}]^{1/2}\right).\label{btoadelta}
\end{equation}
At $\delta=0, a=a_{1}$,\begin{equation}
{b\over a}=\sqrt{1 - {\left( \frac{2}%
      {5\,\left( 5 +{\sqrt{5}} \right) }%
      \right) }^{\frac{1}{3}}- \frac{1}{5}{\left( \frac{2}%
      {5\,\left( 5 +{\sqrt{5}} \right) }%
      \right) }^{-\frac{1}{3}}}.\label{b1toa1}\end{equation}
\section{Isopotentials in GRM} In general form the equation for
critical figure can be written down using variables $\lambda$ and
$\mu$\begin{equation}\label{lamufo}
  \frac{GM}{c}\frac{\lambda }{\lambda ^{2}+\mu
^{2}}+\frac{1}{2}\omega ^{2}c^{2}(1+\lambda ^{2})(1-\mu ^{2})=\Phi
_{0},\end{equation} where $\Phi _{0}$ and $a$ are given in
(\ref{criPhi}) and (\ref{eqrad}). \\From (\ref{criPhi}) we have
quadratic equation for $\mu^{2}$ as function of $\lambda$.
Further, in variables $x,z$ using (\ref{xzlamu}) we get parametric
equation of the complex from (with $\lambda$ as parameter). In
result, we present the Fig. 2 the equilibrium critical figures for
several values of the equatorial radius  $a/c$. With decreasing
$a$ (this corresponding to increasing $\omega$ or increasing $c$,
increasing flatteness of the gravitational field) the deviation of
the GRM critical figures from CRM increases and for small values
of $a$ the figures become even concave while CRM figures are
always convex.
\section{Concave figure} Let us find the value $a=a_{2}$ when the
equilibrium figure becomes concave first. Expanding (1,2) in the
series at small $x \ll 1$ and using (7,8,17), and requiring
$dz/dx=0 \mbox{ at } x=0$ we have the algebraic equation of the
7th order for $y\equiv a_{2}/c$
\begin{equation}\label{a27th}
  144 - 1512\,y + 6408\,y^2 - 14184\,y^3 + 17475\,y^4 -
  11553\,y^5 + 3397\,y^6 -171\,y^7=0;\end{equation}
  similarly for corresponding polar radius $b_{2}$ we have again
  algebraic equations of the same order ($t\equiv b_{2}/c$)
\begin{equation}\label{b27th} 81 + 584\,t + 1278\,t^2 +
  1500\,t^3 + 882\,t^4 +195\,t^5 - 45\,t^6 - 19\,t^7=0.\end{equation}
   Numerically, we get $a_{2}/c=4.00358,\,b_{2}/c=2.0376,$
   and $b_{2}/a_{2}=0.508945.$\\
At $a>a_{2}$ the critical figures in GRM are convex, at $a<a_{2}$
the critical figures (and also inner isopotentials near the
external surface!) are concave.
\begin{figure}[tbh]\centerline{
\includegraphics{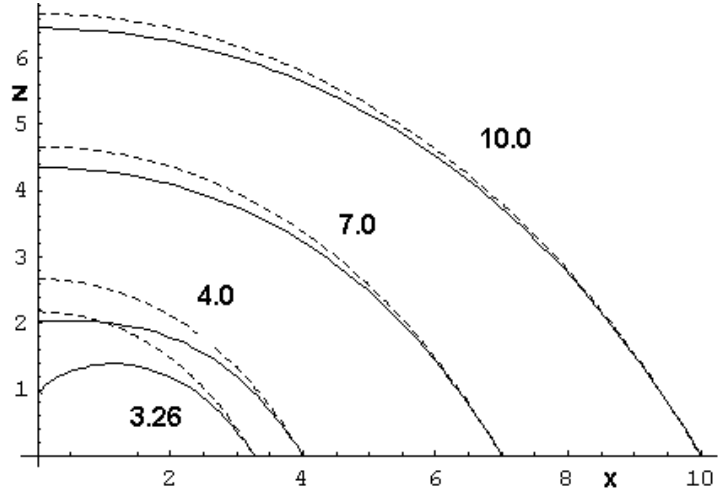}}
\caption{Critical figures in GRM for different values of $a/c$.
    At $a/c >$ 4.0 equilibrium figures are convex, at 3.26$<a/c<$4.0
the equilibrium figures are convex near the polar axis. Also shown
are the equilibrium figures in CGM for the same values of $a$.}
\label{ellipt}\end{figure}
\section{Post-classical approximation}
 The analysis is more easy in
the limit $x\gg c,$ $z\gg c$\bigskip\ (more exactly,
$r=(x^{2}+z^{2})^{1/2}\gg c$). The gravitational potential up to
terms proportional to $c^{2}/r^{2}$ is \begin{equation}
V(r,z)=\frac{GM}{r}\left[1+\frac{1}{2}\frac{c^{2}}{r^{2}}-\frac{3}{2}\frac{%
c^{2} }{r^{2}}\frac{z^{2}}{r^{2}}\right].
\label{vrz}\end{equation} Isopotentials $V(r,z)=V_{c}$
corresponding to (\ref{vrz}) are ellipses with larger semi-axis
$a$ along $x$-axis and minor semi-axis $b$ along $z$-axis
\begin{equation}
a=r_{0}(1+\frac{1}{2}\frac{c^{2}}{r_{0}^{2}}),\ b=r_{0}(1-\frac{c^{2}}{%
r_{0}^{2}}),\ r_{0}=\frac{GM}{V_{c}}.\label{abclass}\end{equation}
At $c=0$ we have the equilibrium figure of the CRM
\begin{equation} r=\frac{G\ M}{\Phi _{c}-\frac{1}{2}\omega
^{2}x^{2}},\label{eqfigclass}\end{equation} or \begin{equation}
r_{0}(x)=\frac{2 a}{3-x^{2}/a^{2}},\,\,0\leq x\leq a,\,\, a\geq
r_{0}\geq b.\label{oreqfigclass}\end{equation} If $\Phi
_{c}>>\frac{1}{2}\omega ^{2}x^{2}$ (slow rotation), the
equilibrium figure is ellipse \begin{equation} r=\frac{G\ M}{\Phi
_{c}}(1+\frac{1}{2}\frac{\omega ^{2}x^{2}}{\Phi
_{c}})\label{ellclass} \end{equation} with larger semi-axes along
the $x$-axis (equatorial radius) and minor semi-axis along the
$z$-axis (polar radius). \\Taking into account terms up to
$c^2/a_0^2$ critical figure is
\begin{equation}
  r(x)=r_{0}(x)(1+\beta c^{2}/a_{0}^{2}),\,0\leq x\leq
 a,\,b\leq r\leq a,
\end{equation}
\begin{equation}
  \beta =-\frac{1}{2t}+\frac{9}{2}t^{4}-3t^{5}-t^{2},\,t\equiv
a_{0}/r_{0},\end{equation} where $r_{0}(x),\,a_{0},a,\,\mbox{and }
b$ are given in Eqs (21-23, 34).
\\Using (36,37) we may calculate the volume of the critical figure
\begin{equation}
  W=4\,\pi\,\int_{0}^{a}\,z(x)\,x\,dx,\,\, z=\sqrt{r^2 - x^2}.
\end{equation}
In (38) both the integrand $z(x)$ and the upper limit $a$ can be
expand in the series up to terms $c^2/a_0^2$

\begin{equation}
  W=4\,\pi\,\left[ \int_{0}^{a_0}\,z_0(x)\,x\,dx,\,+z(a_0)\cdot
a_0\cdot (a-a_0)+\int_{0}^{a}\,z_1(x)\,x\,dx \right],
\end{equation}
$$z_0^2=r_0^2-x^2,\,z=z_0+z_1\,c^2/a_0^2,\,z_1=\beta r_0^2/z_0.$$
Here the first term in W is the volume of critical figure in CRM,
second term is proportional to $c^3/a_0^3$ and may me omitted, and
third term is proportional to $c^2/a_0^2$. \\Both integrals can be
expressed in terms of elementary functions, and we have
\begin{equation}\label{W0}
  \int_{0}^{a_0}\,z_0(x)\,x\,dx=
a_{0}^{3}\int_{0}^{1}\frac{1-x^{2}}{3-x^{2}}(4-x^{2})^{1/2}\ x\ dx=\sqrt{3}-%
\frac{4}{3}+\ln \left( 6-3\sqrt{3}\right),
\end{equation}
\begin{equation}\label{W1}
  \int_{0}^{a_{0}}z_{1}(x)\ x\ dx= \frac{16 - 27\,{\sqrt{3}}}
  {210}a_0^3(c^2/a_0^2).
\end{equation}
  The volume of the critical figure is
  \begin{equation}\label{Wnum}
  W=4\,\pi\,a_0^3\,\left[0.180372-0.146502(c^2/a_0^2)\right].
\end{equation}
  By introducing the mean density of the critical figure
  $\overline{\varrho}=M/W$, we have in the post-classical approximation
for parameter widely used in the theory of rotating configurations
 \begin{equation}\label{param}
  {w^2\over 2 \pi G \overline{\varrho}}={W\over 2 \pi
a_0^3}=0.360744-0.293003\,c^2/a_0^2.
\end{equation}
\section{Conclusion}
We obtained all parameters of rotating configuration in the Roche
model taking into account the effects of (small) non-sphericity of
the gravitational field of the central body. As a rule any
modification of the classical models leads at best to ugly
non-elegant analytical or even pure numerical results.\\ And Roche
models are of the best examples of those classical models
idealized and elegant both in set up and results.
\section{References}
1. Kopal Z. Figures of Equilibrium of Celestial Bodies. Madison:
Univ. of Wisconsin Press, 1960.\\ 2. Tassoul J-T. Theory of
rotating stars. New Jersey: Princeton Univ. Press, 1978.\\ 3. Z.F.
Seidov, 1981DoAze..37...18S;$\,\,$ 1982DoAze..38...30S\\ 4.
Abramowitz, M. and Stegun, I. A. (Eds). "Definition of Elliptical
Coordinates." \S 21.1 in Handbook of Mathematical Functions with
Formulas, Graphs, and Mathematical Tables, 9th printing. New York:
Dover, p. 752,
1972. The book is available also in: http://jove.prohosting.com/~skripty/%
\\ 5. E. Wassermann,
http://mathworld.wolfram.com/ConfocalEllipsoidalCoordinates.html
\end{document}